\documentclass[12pt,preprint]{aastex}

\slugcomment{Draft}

\shorttitle{Charge States in the Fast Solar Wind}
\shortauthors{Laming}

\begin{document}

\title{Ion Charge States in the Fast Solar Wind: New Data Analysis and
Theoretical Refinements}


\author{J. Martin Laming\altaffilmark{1} \& Susan T. Lepri\altaffilmark{2}}


\altaffiltext{1}{E. O. Hulburt Center for Space Research, Naval
Research Laboratory, Code 7674L, Washington DC 20375-5321
\email{laming@nrl.navy.mil}}
\altaffiltext{2}{Department of
Atmospheric, Oceanic and Space Sciences, University of Michigan, Ann
Arbor, MI 48109-2143\email{slepri@umich.edu}}

\begin{abstract}
We present a further investigation into the increased ionization observed in element
charge states in the fast solar wind compared to its coronal hole source regions. Once
ions begin to be perpendicularly heated by ion cyclotron waves and execute large
gyro-orbits, density gradients in the flow can excite lower hybrid waves that then damp
by heating electrons in the parallel direction. We give further analysis of charge state
data from polar coronal holes at solar minimum and maximum, and also from equatorial
coronal holes. We also consider further the damping of lower hybrid waves by ions and the
effect of non-Maxwellian electron distribution functions on the degree of increased
ionization, both of which appear to be negligible for the solar wind case considered
here. We also suggest that the density gradients required to heat electrons sufficiently to further
ionize the solar wind can plausibly result from the turbulent cascade of MHD waves.
\end{abstract}



\section{Introduction}
The mechanism(s) by which electrons and ions can exchange energy collisionlessly, that is in the
absence of Coulomb collisional coupling, have become important issues in a variety of astrophysical
contexts. The fast solar wind provides an excellent laboratory to study such mechanisms in detail due
to the relatively large amount data available.
During its initial polar passes,
in-situ observations from the Solar Wind Ion Composition Spectrometer (SWICS) on {\it Ulysses} revealed that
the fast solar wind is more highly ionized than would be expected given the electron temperature and
charge states spectroscopically observed in its coronal hole source regions.
Ulysses SWICS observed charge states related to electron temperatures near $1.5\times 10^6$ K, assuming
collisional ionization equilibrium \citep{ko97}, contradicting the finding from SUMER on SOHO
which indicate electron temperatures near $8\times 10^5$ K out to a heliocentric distance of
1.5 $R_{\sun}$ \citep{wilhelm98}.  For Fe, temperatures of $8\times 10^5$ K
in the freeze-in region would result in a main charge state of 8+
\citep[see e.g.][]{mazzotta98,bryans06}, which is rarely observed in
any substantial abundance in the solar wind.  Charge states in the fast solar wind for Fe are typically
near 10+.  Hence, any heating mechanism for the electrons between ~ 1.5 solar radii and 3 solar radii
needs to explain the discrepancy between these SUMER observations and the Ulysses SWICS data.

The fast solar wind is likely accelerated
by ion cyclotron waves which damp by heating ions to high perpendicular temperatures
\citep[e.g.][]{cranmer99b,cranmer00}. In an initial
analysis, \citet{laming04} pursued the idea that some of this energy ``leaks'' to the electrons in the
solar wind acceleration region by a collisionless process. The increased electron temperature then
further ionizes the plasma. The most likely collisionless mechanism appeared to be the excitation of
lower-hybrid waves by ions (most likely $\alpha$-particles) executing large gyro-orbits in a
cross-field density gradient. In this way, the electrons are not heated until the ions develop large
gyroradii, beginning at about 1.5 $R_{\sun}$ heliocentric distance. Thus the observational constraint
of \citet{wilhelm98} of electron temperatures of around $8\times 10^5$ K out to 1.5 $R_{\sun}$ is not
violated, which might not be the case for other means of electron heating.

While the existence of such density gradients is highly plausible from Interplanetary Scintillation
observations \citep[see discussion in][]{laming04}, the origin of these structures was not
addressed. The turbulent cascade of MHD Alfv\'en waves excited at the solar surface is known to be essentially
2D in character, proceeding much faster in the direction perpendicular to the magnetic field than
in the direction parallel to it. This means that direct damping of the Alfv\'enic turbulence on ions is
very slow, because the Doppler shift, $k_{\Vert}v_{\Vert}$, is small and few ions are resonant. A possible
solution to this lies in the inclusion of fast mode waves in the turbulence \citep{chandran05,luo06}. In such
cases, the cascade is not confined to the perpendicular direction and high wave frequencies are
accessible. Fast modes are also compressible, and more likely to produce the density gradients observed in
Interplanetary Scintillation, and also invoked in this work to provide the instability that generates lower
hybrid waves. In another approach, \citep{markovskii01,markovskii02,markovskii06} suggest that the
perpendicular gradients of density, magnetic field or velocity produced by the turbulent cascade then
generates cyclotron resonant waves by a plasma instability, which provide the minor ion heating.
We argue that a similar plasma instability should produce
lower hybrid waves that damp on the electrons in the range beyond 1.5 $R_{\sun}$.
\citet{markovskii06} consider in most detail the velocity
shear produced at the high $k_{\perp}$ end of the turbulent cascade, this being the most effective
means of generating ion cyclotron waves.  Hence aside from being
an interesting curiosity with interesting implications in various fields of astrophysics,
our interpretation of the increased ionization of the fast solar wind in terms of small scale perturbations,
in our case density gradients, may have an important bearing
on the central problem of the acceleration of the fast solar wind.

In this paper we provide further analysis of {\it in situ} solar wind ionic composition
data from {\it Ulysses}, for fast wind
from polar coronal holes, and the Advanced Composition Explorer (ACE) for low latitude fast wind streams.
These measurements generally support the earlier results, though the degree by which the ionization
is enhanced appears to be lower than previously reported. We also describe some refinements to the theory,
which combined with the lower observed enhancement in ionization, make it plausible that the cross field
density gradient required is the end result of the MHD turbulent cascade.
This paper is organized as follows. Sections 2 and 3 give some refinements to the theory presented in
\citet{laming04}. Section 2 considers ion heating by lower hybrid waves, and show that this is always
negligible compared to the electron damping of the wave for solar wind conditions. Section 3 discusses
the electron distribution function arising from heating by lower hybrid waves, and shows both on
theoretical and observational grounds that a Maxwellian is the most likely distribution. Section 4 presents
new measurements of the Fe charge states in fast solar wind from polar coronal holes observed
by {\it Ulysses} and equatorial coronal holes observed by ACE, section 5 gives some further discussion and
section 6 concludes.

\section{Ion Heating by Lower Hybrid Waves}
In the model of \citet{laming04}, lower hybrid waves are excited by $\alpha$-particle executing
large gyro-orbits in the presence of a cross field density gradient. These waves are electrostatic
ion oscillations with wavevectors concentrated in a narrow range
(typically $-\omega _{pi}/\omega _{pe}\le \cos\theta\ \le\omega _{pi}/\omega _{pe}$, where $\omega _{pe}$
and $\omega _{pi}$ are the electron and ion plasma frequencies respectively) around the direction
perpendicular to the magnetic field. Electron screening, that would normally inhibit the wave motion,
is restricted for wavelengths greater than the electron gyroradius ($k_{\perp}< v_e/\Omega _e$). Since
$\omega /k_{\perp} >>\omega /k_{\Vert}$, the wave can be simultaneously in Landau resonance with
ions moving perpendicularly to the magnetic field, and electrons moving parallel. Hence a wave generated
by a minor ion executing a large gyro-orbit may in principle be
damped by ions (most likely by protons) as well as by electrons. Ions may Landau damp the wave,
but above a critical wave electric field, $E=B\left(\Omega _i/\omega\right)^{1/3}\omega/4k_{\perp}c$
\citep{karney78}, stochastic heating also sets in. We estimate the heating rates for ions and electrons
as follows. The velocity diffusion coefficient for unmagnetized ions in electrostatic waves
is \citep[][equation 6.43]{melrose86}
\begin{equation}
D_0=\int {8\pi ^2q^2RN\left(\vec{k}\right)\over\hbar\omega}
\left|\vec{e}\cdot\vec{v}\right|^2\delta\left(
\omega -\vec{k}\cdot\vec{v}\right){\hbar ^2k^2\over m^2}{d^3\vec{k}\over\left( 2\pi\right)^3},
\end{equation}
where $R$ is the ratio of electric to total energy in the wave, $N\left(\vec{k}\right)$ is the
number density of wave quanta with wavevector $\vec{k}$, $\vec{e}$ is the wave polarization vector
and $\vec{v}$ is the ion velocity. Writing $E^2/8\pi =\int RN\left(\vec{k}\right)\hbar\omega
d\vec{k}^3/\left(2\pi\right)^3$ we obtain
\begin{equation}
D_0={\pi q^2E^2k^2\over m^2\omega ^2}v^2\cos ^2\psi\delta\left(\omega -kv\cos\psi\right)=
{\pi q^2E^2\over m^2}\delta\left(\omega -kv\cos\psi\right).
\end{equation}
Following \citet{karney79},
the diffusion coefficient for magnetized ions in perpendicularly propagating electrostatic waves
is obtained from this result by averaging over Larmor angle $\psi$, with $\vec{k}\rightarrow
k_{\perp}$, $\vec{v}\rightarrow v_{\perp}$,
\begin{equation}
D_B={1\over 2\pi}\int _0^{2\pi}\cos ^2\psi {\pi q^2E^2\over m^2}\delta\left(\omega -k_{\perp}
v_{\perp}\cos\psi\right)d\psi= {q^2E^2\over m^2}{\omega ^2\over k_{\perp}^2v_{\perp}^2}
{1\over k_{\perp}v_{\perp}\sqrt{1-\omega ^2/k_{\perp}^2v_{\perp}^2}},
\end{equation}
where we have assumed two resonances per gyro-orbit, i.e. the $\delta$ function is satisfied
twice in $2\pi$. The heating rate is
\begin{equation}
Q={m_i\over 2}\int v_{\perp}^2{\partial f_i\over\partial t}d^3\vec{v}=-2\pi m_i\int v_{\perp}^2
D{\partial f_i\over\partial v_{\perp}}dv_{\perp}dv_{\Vert}
\end{equation}
where the second step follows from writing $\partial f/\partial t=\left(1/v_{\perp}\right)
\partial /\partial v_{\perp}\left(v_{\perp}D\partial f_i/\partial v_{\perp}\right)$, and
integrating by parts in cylindrical coordinates. Substituting the zero field ion velocity diffusion
coefficient, the heating rate
\begin{equation}Q={\pi q^2E^2\over m}\int {v_{\perp}^2\over v_{t\perp}^2}f\delta\left(\omega -
\vec{k}\cdot\vec{v}\right)2\pi v_{\perp}dv_{\perp}dv_{\Vert}={2\sqrt{\pi }\omega W\over n_i}
\left(\omega\over\sqrt{2}k_{\perp}v_{\perp}\right)^3\exp\left(-{\omega ^2\over 2k_{\perp}^2
v_{\perp}^2}\right),
\end{equation}
where we have put the wave energy density $W=\omega\left|E_0\right|^2 /8\pi\times\partial\epsilon _L/
\partial\omega = \omega _{pi}^2\left|E_0\right|^2/4\pi\omega ^2$.
The final step may be rewritten $Q=2W\gamma _{LD}/ n_i$, where
$\gamma _{LD}=\sqrt{\pi}\omega x^3\exp -x^2$ with $x=\omega /\sqrt{2}k_{\perp}v_{t\perp}$
is the ion Landau damping rate \citep[see e.g.][equation A12 with $U=0$]{laming01}.

In nonzero magnetic field, we substitute equation 3 into equation 4, and with
$u^2=v_{\perp}^2/2v_{t\perp}^2$, we get
\begin{equation}
Q={2W\over n_i}\omega x^3\int _x^{\infty}{u\over\sqrt{u^2-x^2}}\exp -u^2du={w\over n_i}
\omega x^3\int _{x^2}^{\infty}{\sqrt{v}\over\sqrt{v-x^2}}\exp -vdv
\end{equation}
with $v=u^2$. The last form of the integral in equation 6 may be evaluated in terms of a
Whittaker function \citep[][3.383.4, p319]{gradshteyn94} to give
\begin{equation}
Q={W\over n_i}\omega x^3\Gamma\left(1\over 2\right)\exp\left(-{x^2\over 2}\right)
W_{1/2,-1/2}\left(x^2\right).
\end{equation}
This is then evaluated using \citet{abr_st} equation 13.1.33 (p190) to give
\begin{equation}
Q={W\over n_i}\omega x^3\sqrt{\pi}\exp -x^2U\left(-{1\over 2},0,x^2\right).
\end{equation}
Using limiting forms for the confluent hypergeometric function $U\left(-1/2,0,x^2\right)$
\citep[][equations 13.5.11, p193 and 13.5.2, p193]{abr_st} we finally arrive at
\begin{equation}
Q=\cases {\left(W/n_i\right)\omega x^3\exp -x^2 &$x^2\rightarrow 0$;\cr
\left(W/n_i\right)\omega x^4\sqrt{\pi}\exp -x^2 &$x^2\rightarrow\infty$\cr}
\end{equation}
Thus in the limit $x>>1$, the magnetized ions are heated a factor of $x/2$ faster
than the unmagnetized ions. \citet{brambilla98} gives a similar result (page 504), but his
diffusion coefficient is different from our equation 3 by a factor $\Omega _i/2k_{\perp}v_{\perp}$,
where $\Omega _i$ is the ion gyrofrequency. For $x\rightarrow 0$, the unmagnetized Landau damping
is a factor $2\sqrt{\pi}$ faster.

The electron Landau damping rate is given by equation A11 in \citet{laming01} as
\begin{equation}
2\gamma ={\pi\omega ^2\over k^2n_i}\left[1+{\omega _{pe}^2\over\omega _{pi}^2}{k_{\Vert}^2
\over k_{\perp}^2}\left(1+{\omega _{pe}^2\over k_{\perp}^2c^2}\right)^{-1}\right]^{-1}
{\omega _{pe}^2\over\omega _{pi}^2}\left(1+{\omega _{pe}^2\over k_{\perp}^2c^2}\right)^{-1}
{\partial f_e\over\partial v_{e\Vert}}\left(v_{e\Vert}=\omega /k_{\Vert}\right)
\end{equation}
which gives a damping rate of order $\sqrt{\pi}\omega m_i/m_e y^3\exp -y^2/2$ where
$y=\omega/\sqrt{2}k_{\Vert}v_{e\Vert}$. For $x\sim y$, this is $m_i/m_e$ faster than ion
Landau damping, and approximately $\sqrt{m_i/m_e}$ faster than the stochastic heating rate
for magnetized ions. We conclude that lower hybrid waves damp almost exclusively by heating
electrons, and that the electric field in the wave may be higher than that set by \citet{karney78}.
For ion heating to be significant, we require $y >> x >> 1$, i.e. the ions should already be significantly
hotter than the electrons, but still in a ``cold plasma'' regime. The only qualification to this might
be if the lower hybrid waves propagate at cosines $\cos ^2\theta << m_e/m_i$ so that $k_{\Vert}\rightarrow 0$.
This seems unlikely, since even in the cold plasma case where lower hybrid growth rates and
frequencies remain nonzero at $\cos ^2\theta =0$, they are still lower than at $\cos ^2\theta =m_e/m_i$, and
so the existence of waves with nonzero $k_{\Vert}$ and the consequent electron heating
would seem to be inevitable.

\section{Electron $\kappa$ Distributions in the Fast Solar Wind?}
Electrons heated by collisionless processes do not necessarily maintain a Maxwellian
velocity distribution function. In this section we investigate the effect of a nonMaxwellian
electron velocity distribution function on the ionization
balance of oxygen. The electrons are taken to be be heated into a $\kappa$ distribution,
$f\left(v\right)=\left(1+v^2/2\kappa v_t^2\right)^{-\kappa}$. As $\kappa\rightarrow\infty$,
$f\left(v\right)\rightarrow\exp\left(-v^2/2v_t^2\right)$ and a Maxwellian is recovered.
In Appendix B we give a simple derivation of this form for the electron velocity distribution
function from lower hybrid wave heating. In the case that $\omega <<\sqrt{2}kv_i$, a $\kappa$
function results, with the value of $\kappa$ dependent on the wave electric field and the ambient
density. In the opposite limit, $\omega >>\sqrt{2}kv_i$, a Maxwellian results, with
quasi-thermal velocity also dependent on the wave electric field. Based on our knowledge of
coronal parameters, we would expect the latter case to be most likely for the acceleration
region of the solar wind, where the ions are further ionized. In this section we show that the
electrons in a Maxwellian distribution function are also more consistent with the observed
charge states than are $\kappa$ distributions.

We assume that the electron $\kappa$ distribution is isotropic, mainly for ease of computation.
We have
\begin{eqnarray}
&\int _0^{\infty}4\pi v^2\left(1+{v^2\over 2\kappa v_t^2}\right)^{-\kappa}dv=\left(2\pi v_t^2\kappa\right)
^{3/2}{\Gamma\left(\kappa -3/2\right)\over\Gamma\left(\kappa\right)}=n\cr
&\int _0^{\infty}{1\over 2}mv^24\pi v^2\left(1+{v^2\over 2\kappa v_t^2}\right)^{-\kappa}dv=
{3m\over 4\pi }\left(2\pi v_t^2\kappa\right)^{5/2}{\Gamma\left(\kappa -5/2\right)\over
\Gamma\left(\kappa\right)}={3\over 2}nkT,
\end{eqnarray}
so $v_t^2=\left(1-5/2\kappa\right)kT/m$. The energy in the ``tail'' portion of the distribution
function is $\left(3/2\right)\left(nkT-nmv_t^2\right)=15nkT/4\kappa$. We write the temperature
equilibration rate between the ``core'' and the ``tail'' as $5T/2\kappa\tau _s$ where
$\tau _s=2\pi nv^3/3\omega _{pe}^4\ln\Lambda$ is the stopping time by Coulomb collisions for
fast electrons in cold plasma.

\begin{figure}
\epsscale{0.5} \plotone{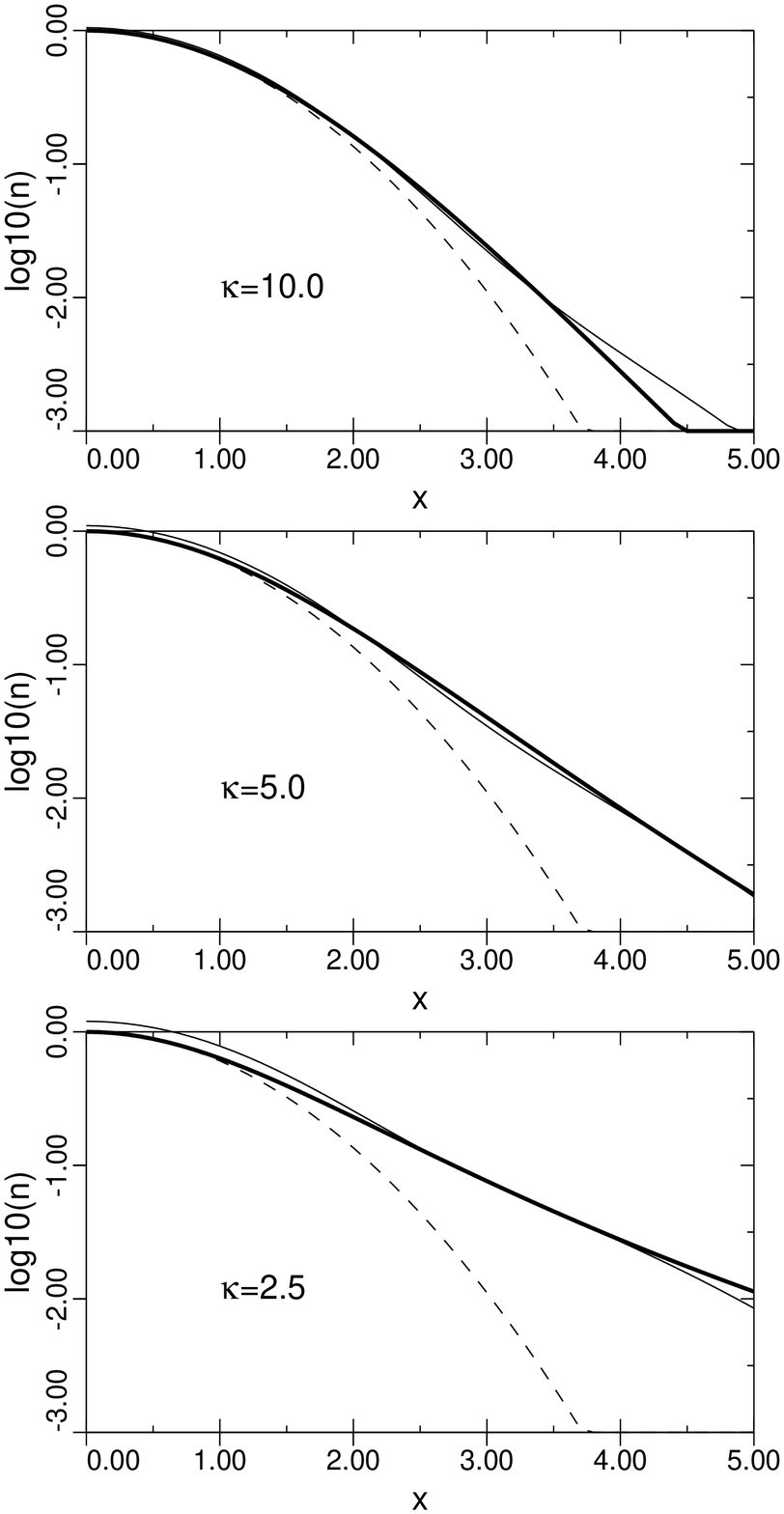}
\figcaption[f1.eps]{Approximate (narrow solid line, from equation 12) and true (wide solid line)
$\kappa$
distributions for varying $\kappa$. The underlying Maxwellian is shown in each case as a dashed line.
\label{fig1}}
\end{figure}

We modify the ionization and recombination rates according to the departure from the Maxwellian
distribution function. In principle, the cross section for each process should be integrated over
the $\kappa$ distribution to evaluate the new rate. This is straightforward for direct collisional
ionization and radiative recombination, but for the resonant processes of dielectronic recombination
and autoionization following inner shell excitation, becomes considerably more cumbersome. Instead, we
approximate the $\kappa$ distribution by
\begin{equation}
f\left(v\right)=\left(1+v^2/2\kappa v_t^2\right)^{-\kappa}=\exp\left(-v^2\over 2v_t^2\right)+
{1\over 2\kappa }\exp\left(-v^2\over 6v_t^2\left(1+5/\kappa ^3\right)\right)+ ...
\end{equation}
which has been found by trial and error. We illustrate the comparison between the true and
approximate $\kappa$ distributions in Figure \ref{fig1}. We then evaluate approximate rate
coefficients by making the appropriate sum over Maxwellian rate coefficients, and proceed with
simulations in the manner described in \citet{laming04}. We keep the same definition of the
electron-ion equilibration parameter, $\gamma ^{\prime}=\gamma _iM_i/\omega Afq^2\times\left(\omega ^2/k^2
v_{\perp}\right)$ where $\gamma _i$ is the lower hybrid wave growth rate due to the gyrating ions,
$M_i$ is the ion mass, $A$ is the ion abundance, $f$ the charge state fractions, $q$ the charge,
$\omega$ the lower hybrid wave frequency, $k$ the wavevector and $v_{\perp}$ the ion gyration velocity.
The ion-electron energy transfer rate is modified by a factor $\times 2\omega _{pe}^2/\Omega _e^2$. This
corrects an error in \citet{laming04} where only the electric energy density of the lower hybrid wave
was considered in computing the energy transfer as $2W\gamma _e$, whereas $W$ should properly be
the total wave energy density.
Thus for a given $\gamma ^{\prime}$, the electron internal energy, $3nkT/2$, is the same in these models as
in those of \citet{laming04}, multiplied by this correction factor of $2\omega _{pe}^2/\Omega _e^2\simeq
20$. Each simulation is initiated with a high value of $\kappa = 10^3$ to enforce an initial Maxwellian
distribution close to the sun.
After every time step in which the electron and ion temperatures and charge states are integrated, we
update $v_t^2$ along with the other hydrodynamic variables in operator splitting fashion, according to
\begin{equation}
{dv_t^2\over dt}={5\over 2\kappa}{kT\over m\tau _s}
\end{equation}
and then update $\kappa = 2.5/\left(1-mv_t^2/kT\right)$. This assumes that all of the electron
heating goes to electrons in the tail. The Coulomb equilibration between the ``tail'
and the ``core'' is slower than the overall collisionless heating rate, and so $\kappa$ decreases
as the heating proceeds.


Table 1 gives values of the O$^{+6}$/O$^{+7}$ charge state ratio for models with
varying $\gamma ^{\prime}$, for initial wind flow speeds of 5, 10, and 20 km s$^{-1}$. Charge states
are given for two cases, one with $\kappa$ evaluated according to the procedure given above, and
one with $\kappa\rightarrow\infty $, i.e. a Maxwellian distribution is enforced throughout as in
\citet{laming04}. Except in cases with the lowest values of electron-ion equilibration, a $\kappa$
electron velocity distribution produces {\em less} ionization than the Maxwellian distribution. With
little or no electron-ion equilibration, the $\kappa$ distribution has more electrons above the O$^{+6}$
ionization threshold than the corresponding Maxwellian, and so higher ionization states result. At
higher electron temperatures, many electrons are above the threshold in the Maxwellian, and the
$\kappa$ distribution puts more of these at even higher energies, where the ionization cross section
is starting to decrease. Thus at high electron-ion equilibration rates, the $\kappa$ distribution
produces less ionization. We have previously identified these cases as being necessary to
produce the observed degree of ionization, and so we conclude that at least in the region of the
fast solar wind interior to freeze-in, the electrons are heated by lower hybrid waves into a Maxwellian
distribution function, as would be expected in the case that $\omega >>\sqrt{2}kv_i$ (see Appendix B).
Henceforward, $\kappa\rightarrow\infty $ is assumed in all models. The electron power law component
observed by Ulysses \citep[e.g.][]{maksimovic00} must be generated by another heating mechanism,
possibly by lower hybrid waves
further out in the solar wind where the magnetic field is reduced so that $\omega <<\sqrt{2}kv_i$, or by
the damping of whistler turbulence \citep[e.g.][]{vocks05}. This conclusion is at variance with the
assumption of \citet{esser00}, who argued that such a tail on the electron distribution should
persist all the way down to the Sun, based on the insensitivity to such an electron
distribution of the electron temperature diagnostic used by \citet{wilhelm98}. However, as discussed
previously \citep{laming04}, other diagnostics exist which are sensitive to suprathermal electrons,
and these generally reveal insignificant populations. Such suprathermal electron tails are often
attributed to the Landau damping of kinetic Alfv\'en waves \citep{vinas00}, which if true, provides
no obvious reason for the electron heating to be suppressed below 1.5 $R_{sun}$. By coupling the electron
heating to the ion acceleration as in our model, this constraint is automatically satisfied.

\section{Fe Charge States Measured by Ulysses and ACE in Polar Fast Wind and the Equatorial Fast Wind}

The creation and destruction of an ion species is governed by the following rate equation for its density
$n_i$,
\begin{equation}
{dn_i\over dt} = -C_i n_i n_e - (R_{dr,i} + R_{rr,i})n_i n_e + C_{i-1}n_{i-1}n_e + (R_{dr,i+1} +
R_{rr,i+1})n_{i+1}n_e - 2n_iv/r
\end{equation}
where $n_e$ is the electron density, $C_i$ is the collisional ionization rate,
$R_{dr,i}$ is the dielectric recombination rate, $R_{rr,i}$ is the radiative recombination rate, and $v$
and $r$ are the solar wind velocity and radius.
These the ionization and recombination rates depend on the local electron temperature and the speed of the
ion. When the electron density is reduced by the expansion such that the last term dominates on the
right hand side, the charge states are said to have frozen-in. Depending on the element concerned,
freeze-in occurs between 1.5 and 3.5 $R_{\sun}$ heliocentric distance.
Observation of in-situ charge states therefore allows the inner corona to be studied in detail.

We examined charge state distributions of Fe for a variety of fast solar wind periods using data we
obtained from both ACE and Ulysses.  The Ulysses SWICS data has a time resolution of 3 hrs  (obtained
via private communication, R. von Steiger), while the ACE SWICS data has a time resolution of 1 hr
(obtained from the ACE Science Center).  The coronal holes were identified by examining the solar
wind speed, the O$^{+7}$/O$^{+6}$ ratio, and the C$^{+6}$/C$^{+4}$ ratio.  The mean charge state
distributions were calculated for the 2 $-$ 4 day period when either Ulysses or ACE was in a high
speed stream not attributed to ICMEs, but consistent with coronal hole solar wind. We examine
three different types or cases of fast solar wind; (1) four periods observed in a single
well-formed solar minimum polar coronal hole, (2) four periods inside a single solar maximum
polar coronal hole, and (3) 14 periods inside separate fast solar wind streams from solar
maximum equatorial coronal holes observed at ACE in 2005. We will compare the observed charge
state distributions with those predicted using various heating rates.

Figure 2 shows Fe charge states observed in the fast solar wind by Ulysses (Case 1).  There are
six time periods, four days in duration, identified as wind originating in the polar coronal holes
during the southern polar pass.  The relative abundance of the charge states for these four day
periods are averaged and plotted. The distributions tend to peak between $Q_{Fe}$ = 10 - 11.
This is consistent  with the finding of \citet{ko99} for the years 1994 - 1995.  A solution
from the model with $\gamma ^{\prime}$ = 0.025 and an initial speed at $1.05 R_{\sun}$ of
10 km s$^{-1}$ \citep[the preferred value from][]{laming04}
is shown to closely match the observed distribution function.
Fe charge states for a variety of these models are given in Table 2, and $\gamma ^{\prime}$ is
defined in section 3.
This demonstrates how lower hybrid wave damping can additionally heat the solar wind before
the charge states freeze-in matching observations, and that only one characteristic density scale
length is required.

Figure 3 shows the Fe charge state distributions for a solar maximum coronal hole observed by
Ulysses SWICS.  These four four-day periods occurred during the northern polar pass toward the end
of 2001.  Again the charge states represent the mean charge state distribution observed during these
periods.  The maximum charge state for this period is near 10, although a large fraction remains
in $Q_{Fe}$ = 9. These distributions are compared to a model with
a heating rate of $\gamma ^{\prime}$ = 0.015. However it is clear that the observed charge state
distributions are broader than the model, indicating that a range of density scale lengths may be
present.

Figures 4 and 5 show the Fe charge state distribution for fast solar wind associated with an equatorial
coronal hole observed by ACE SWICS during solar minimum.  For these plots, 14 periods are selected
ranging from 2-4 days in duration.  These periods represent different passes through fast solar
wind, whereas the previous two examples represent different periods within the same coronal hole.
The periods are separated into those that peak at $Q_{Fe}$ = 9  (Figure 4) and those that peak
near $Q_{Fe}$ = 10  (Figure 5).  The charge state distributions that peak at $Q_{Fe}$ = 9 are fit
best by a model $\gamma ^{\prime}$  = 0.01, while the charge state distributions that peak at
$Q_{Fe}$ = 10 are best fit by a model $\gamma ^{\prime}$  = 0.015, which implies more heating
by wave damping is needed for these.  The model results fail to catch the shoulder of the charge
state distributions out to a charge state of 14 and beyond.

Since the ionization model assumes solar minimum conditions over the poles, there may be additional
characteristics that need to be considered for a full understanding of the contribution of wave
damping for the heating of ions in different solar wind streams from coronal holes.
\citet{miralles01} observed that the O$^{5+}$ outflow speed was lower for the equatorial coronal
holes than for the polar coronal holes, despite the fact that their final solar wind speeds at 1 AU
are not significantly different. The wind acceleration must occur further out, possibly beyond
the radii associated with freeze-in, leading to lower degrees of enhanced ionization. It may be that
the higher densities in these structures prevent the plasma from becoming collisionless further
from the solar surface than is the case in polar coronal holes, or that the increased magnetic
field strength \citep{wang00} inhibits the turbulent cascade until the field has decreased
sufficiently further out, since three-wave interaction probabilities go as $\left(\omega /
\Omega _i\right)^4$ \citep{luo06}.

Ulysses charge states from the solar minimum polar coronal hole during 1994 are observed to be higher
than the charge states in the solar maximum polar coronal hole observed by Ulysses in 2001.  The
charge states in the equatorial coronal hole observed at ACE in 2005 are even lower, with an average
charge state of 9+ for Fe.  The trend toward lower average charge states correlates with the average
of the solar wind speeds during the periods measured.  The average speed of the solar wind alpha
particles observed from the solar minimum equatorial coronal hole at ACE was 650 km/s (ranging
from 600 - 800 km/s) , while solar wind alpha speeds from the solar minimum polar coronal hole at
Ulysses were at 800 km/s +/- 20 km/s.  The wind speeds were more variable in the ACE observations.
The relationship of solar wind alpha speed and mean Fe charge state and therefore lower hybrid
damping rate indicates that less additional electron heating is needed in slower flows at lower latitudes.
The electron temperatures required to produce the observed charge are slightly lower than was the
case in \citet{laming04}, due to the small downwards revision of these charge states. We find the
collisionless equilibration parameter $\gamma ^{\prime}$ in the range 0.01 - 0.025, which corresponds
to the curves labeled 0.2 - 0.5 in Figure 5 of \citet{laming04}. This represents a small change, becoming
a little more consistent with the observational error bars from \citet{wilhelm98}.

\section{Discussion: Relation to MHD Turbulence}
The cascade of Alfv\'enic MHD turbulence from small to large $k_{\perp}$ is expected to produce cross
field variations on small length scales of density, magnetic field and velocity. \citet{markovskii06}
consider in most detail the velocity shear produced by perpendicularly propagating Alfv\'enic
fluctuations. Such perturbations will also produce small scale perpendicular variations in magnetic field.
While magnetic curvature may also give rise to lower hybrid waves, it is the longitudinal variation
that is important in this respect. However $k_{\Vert}$ does not cascade to large values, unlike $k_{\perp}$,
so we are left with cross-field density gradients as the most effective means of generating
lower hybrid waves. Significant density fluctuations cannot arise solely from Alfv\'enic turbulence. Waves
of a compressive character such as fast mode waves are required. \citet{luo06}, in an analytic
consideration of three-wave processes involving two or three Alfv\'en waves and at most one fast mode wave,
show that a critical frequency, $\omega _c$, exists,
such that below this frequency fast mode waves dominate, and above Alfv\'en waves are more prevalent.
In the simplest case involving one fast mode wave and two Alfv\'en waves, with one of the Alfv\'en
wave frequencies, $\omega _l$, much lower than the other two,
$\omega _c\sim 2\omega _{pi}^{2/3}\omega _l^{1/3}$, putting $W_F\sim W_A$ in their equation 25. For
a density of $10^6$ cm$^{-3}$ and magnetic field of 1 G, $\omega _l\sim 0.07$ which corresponds to a
wave period of 90 s, assuming that in our case, fast mode turbulence must dominate up to frequencies
corresponding to the ion gyrofrequency. Waves of this period or longer are reported in coronal holes,
\citep{ofman99,banerjee01}, and so the existence of fast mode turbulence would not seem to be a problem.
\citet{chandran05}, adopting a numerical approach to consider a wider range of processes involving Alfv\'en
and fast mode waves, finds different regions of $k$-space dominated by Alfv\'en or fast mode turbulence.
\citet{suzuki06} conjecture that the degree of coupling between fast modes and Alfv\'en modes
should be reduced in strong Alfv\'enic turbulence; both \citet{luo06} and \citet{chandran05} only
consider weak Alfv\'enic turbulence.
Slow mode waves have not been considered as part of the turbulent cascade, but \citet{ballai03} discuss
the behavior of slow mode solitary waves, which could also produce suitable density gradients in
perpendicular propagation.

Direct evidence of turbulent cascade down to these small scales has been
hard to come by. However Cluster \citep{bale05} has observed electric and magnetic fluctuation spectra
in the solar wind near the terrestrial bowshock. In the inertial range, both spectra resemble a
Kolmogorov $k^{-5/3}$ behavior. However as the dissipation range is approached, at $kr_g\sim 1$ where
$r_g$ is the ion gyroradius, the magnetic spectrum steepens, while the electric spectrum flattens. Clearly
the turbulence is becoming more electrostatic. Measurements of the phase speed suggest the turbulence
is mainly Alfv\'enic in nature, transforming to kinetic Alfv\'en waves in the dissipation range and
damping by heating the ambient plasma as the waves become more electrostatic. This is entirely consistent
with the approach taken in this paper, except that the electrostatic waves we consider derive not
directly from the cascading turbulence, but as a by-product of the damping.

We determine a likely order of magnitude of $\gamma ^{\prime}\sim 0.01$ from charge states observed in the fast
solar wind. Assuming a wave electric field given by equation A3 and a Maxwellian distribution for
the $\alpha$ particles that excite the waves, we determine a density scale length $L\sim 4r_g$ from
figure 11 in \citet{laming04}, where $r_g$ is the gyroradius of the $\alpha$ particles.
Observations of $\alpha$ particles in the solar wind typically give
a $\kappa$ distribution in the parallel direction with $\kappa\sim 3-6$ \citep{collier96,chotoo98},
which if it has an origin in energization by ion cyclotron waves would correspond to $\kappa\sim 3.5-6.5$
in the perpendicular direction. This would increase our density scale length estimate to $L\sim 6-10 r_g$.
We showed earlier that ion heating is negligible compared to electron heating for lower hybrid waves
excited under solar wind parameters. Thus the electric field estimate in equation A3 may be exceeded
by as much as an order of magnitude. Hence $\gamma ^{\prime}\sim 10^{-4}$, and the density scale
length becomes $L\sim 10-25 r_g$. We have no direct knowledge of the $\alpha$ particle gyroradius. The
perpendicular velocities for protons and O VI are $\sim 200$ and $\sim 300-400$ km s$^{-1}$ at
a heliocentric distance of about 2 $R_{\sun}$, where the magnetic field is $\sim 1$ G and the particle
density is $\sim 10^6$ cm$^{-3}$. The proton gyroradius is then 0.02 km, and the proton inertial
length is 0.2 km. At 1 AU, the minor ions flow about an Alfv\'en speed faster than the protons,
presumably reflecting enhanced perpendicular heating by ion cyclotron waves closer to the sun. The
$\alpha$ particles also flow faster than the protons \citep{reisenfeld01,vonsteiger00,neugebauer96}, so we
will assume an $\alpha$ particle perpendicular velocity of 400 km s$^{-1}$, similar to O VI. This gives
an $\alpha$ particle gyroradius of 0.08 km, and the range of density scale lengths corresponds to
0.8 - 2 km, or 4 - 10 proton inertial lengths. This is very similar to the scale length of velocity
shear discussed by \citet{markovskii06}. Thus it seems that our results support ideas of turbulent
cascade of MHD fluctuations in the solar wind. Other observations of density structures observed
by Interplanetary Scintillation are reviewed in \citet{laming04}.

Another often cited mechanism for electron heating in the solar wind is the Landau damping of
kinetic Alfv\'en waves \citep[e.g.][]{cranmer03} by the small parallel component of the wave
electric field. This process, and that for fast mode waves also, are considered in some detail by
\citet{vinas00}, with application to the formation of suprathermal electrons in the chromosphere.
Their solution of the dispersion relation only yields significant parallel electric fields for wavevectors
approaching $10 c/\omega _{pi}$ for both fast mode and kinetic Alfv\'en waves. At $k\sim c/\omega _{pi}$,
the parallel electric field is a factor $\sim 10^{-4}$ smaller than the perpendicular field, and smaller
by at least an order of magnitude than the parallel electric field due to lower hybrid waves if
the density gradients we infer above are related to Alfv\'en or fast mode wavevectors. More
generally, one can estimate the parallel electric field from the expression \citep[][equation 37]{hasegawa76}
\begin{equation}
E_{\Vert}=r_g^2{T_e\over T_i}\partial _{\Vert}\left(\partial _{\perp}E_{\perp}\right),
\end{equation}
which would suggest that wavelengths of order the proton gyroradius are required. This is less restrictive than
the results of \citet{vinas00}, but still significantly more so than what we determine for lower hybrid
waves. It
would appear then that electron heating by kinetic Alfv\'en waves places even more stringent
requirements on the turbulent cascade than do the lower hybrid waves. Another potential problem is that
no natural means of suppressing electron heating at altitudes within 1.5 $R_{\sun}$ heliocentric
distance exists, which might lead to conflict with the SUMER electron temperature observations of
\citet{wilhelm98} and others.

\section{Conclusions}
We have revisited the model of \citet{laming04} for increased ionization in the fast solar wind with
the analysis of new data, and some refinements to the theory. The important new development, due in part
to the revision downwards of charge states observed in the solar wind, and more significantly to the
correction of an error in \citet{laming04} in the expression for the wave energy density, is that the
degree of density inhomogeneity required in the fast solar wind acceleration region to produce the
required heating is now reduced from a length on the order of an ion gyroradius, to be in the range 10 - 25
gyroradii. As discussed above, this makes a connection between these density gradients and the outcome
of the MHD turbulent cascade much more likely. Our finding that ion excited lower hybrid waves
develop stronger parallel electric fields than kinetic Alfv\'en waves reinforces suggestions like
that of \citet{begelman88} that such mechanisms may have more general relevance elsewhere in astrophysics.

In our models, the appropriate choice of the electron-ion equilibration parameter can yield a fit
for the observed distribution functions for various fast solar wind streams.  This supports the idea
that existence of density gradients from MHD turbulent cascades can allow lower hybrid waves to oscillate
and damp, imparting their energy to electron heating.  This electron heating due to the damping of the
lower hybrid waves occurs in the region beyond $1.5 R_{\sun}$, where ions begin to execute large gyro-
orbits under the influence of perpendicular heating by ion cyclotron waves.
The solar wind flows with the highest alpha particle speeds also have the
highest average charge states, implying that the electron heating is most efficient in fast outflow from
polar coronal holes during solar minimum.  Lower fast wind speeds result in lower average charge states of
Fe, clearly indicating a connection between electron heating and ion acceleration. Finally, in Table 3,
we give predicted charge states for Ne, Mg, Si, S, Ar, and Ca, again for an initial wind speed of
10 km s$^{-1}$ at $1.05 R_{\sun}$.

\acknowledgements
This work has been supported by NASA LWS Grant NNH05AA05I (JML) and by an NSF SHINE Postdoctoral
Fellowship ATM-0523998 (STL).

\appendix
\section{Wave Trapping and Threshold Electric Field}
The equation of motion for a particle in the electric field of a wave is
\begin{equation}
{dv\over dt}={qE\over m}\exp i\psi ={qE\over m}\cos\psi
\end{equation}
where $\psi = \omega t-kz$ and real parts are taken in the last
step. Writing $\partial ^2z/\partial t^2=-\left(1/k\right)\partial
^2\psi /\partial t^2=qE/m \cos\psi$ we arrive at
\begin{equation}
{\partial ^2\over\partial t^2}{\partial\psi\over\partial
t}={qEk\over m}{\partial\psi\over\partial t}\sin\psi =-\omega
_T^2{\partial\psi\over\partial t}\sin\psi
\end{equation}
where $\omega _T=\sqrt{-qEk/m}$ is the particle trapping frequency. A particle is
said to be ``trapped'' when $v>\left(\omega -\omega _T\right)/k$ during its cyclotron
orbit. The time spent in the trapping region during a cyclotron orbit is
$2\theta /\Omega$ where $\cos\theta =1-\omega _T/\omega\simeq 1-\theta ^2/2$, so the
time is $2/\Omega \times\sqrt{2\omega _T/\omega }$. The trapping is defined as effective
when the time spent in the trapping region is $\ge 1/\omega _T$, so $8\omega _T^3=\Omega ^2\omega$,
which then gives
\begin{equation}
{E_0\over B_0}={1\over 4}\left(\Omega\over\omega\right)^{1/3}{\omega\over kc},
\end{equation}
which is the result of \citet{karney78}, derived by integration of the Hamiltonian equations of motion.

\section{Electron Distributions from Lower-Hybrid Wave Heating}
We write the Boltzmann equation as
\begin{equation}
{df\over dt}={1\over v_{\perp}}{\partial\over\partial v_{\perp}}\left[
v_{\perp}\left\{A_{\perp}f+\left(D_{\perp\perp}{\partial f\over\partial
v_{\perp}} +D_{\perp\Vert}{\partial f\over\partial v_{\Vert}}\right)\right\}
\right]+{\partial\over\partial v_{\Vert}}\left\{A_{\Vert}f+D_{\Vert\perp}
{\partial f\over\partial v_{\perp}}+D_{\Vert\Vert}{\partial f\over\partial
v_{\Vert}}\right\}=0,
\end{equation}
where $A_{\perp}$ and $A_{\Vert}$ are coefficients of dynamical friction and $D_{\perp\perp}$,
$D_{\Vert\Vert}$, $D_{\perp\Vert}$, and $D_{\Vert\perp}$ are velocity diffusion coefficients.
Integrating the Boltzmann equation over $v_{\perp}$ from 0 to $\infty$, we have
\begin{equation}
{\partial\over\partial v_{\Vert}}\left\{A_{\Vert}\int f2\pi v_{\perp}dv_{\perp}
+\int D_{\Vert\perp}{\partial f\over\partial v_{\perp}}2\pi v_{\perp}dv_{\perp}
+D_{\Vert\Vert}{\partial \over\partial
v_{\Vert}}\int f2\pi v_{\perp}dv_{\perp}\right\}=0
\end{equation}
in steady state conditions (i.e. $df/dt=0$).
This integrates to
\begin{equation}\int f2\pi v_{\perp}dv_{\perp}\propto\exp\left(
-\int A_{\Vert}/D_{\Vert\Vert} dv_{\Vert}\right)
=\exp\left(-v_{\Vert}^2/2v_{t\Vert}^2\right)
\end{equation}
when $A_{\Vert}=v_{\Vert}/\tau$,
$D_{\Vert\perp}=v_{t\Vert}v_{t\perp}/\tau _d$ and
$D_{\Vert\Vert}=v_{t\Vert}^2/\tau _s$, with the deflection time $\tau _d >> \tau _s$, the
stopping time,  $\tau _s = 2\pi n_ev_{\Vert}^3/3\omega _{pe}^4\ln\Lambda$ for fast electrons
in cold plasma, with $\ln\Lambda\sim 30$ being the Coulomb logarithm. For comparison, the
electron-electron equilibration time, $t_{eq}=\sqrt{2/3}\pi ^{3/2}n_ev^3/\omega _{pe}^4\ln\Lambda$, is a
factor of a few higher.

For electron heating by Landau damping, the non-zero diffusion
coefficient and coefficient of dynamical friction are
\citep[][equation 10.83 and 10.90 with $s=0$]{melrose86}
\begin{eqnarray}
D_{\Vert\Vert}&=\int {8\pi ^2q^2\over\hbar}{R\over
\omega }\left|\vec{e}\cdot\vec{V}\right|^2\delta\left(\omega
-k_{\Vert}v_{\Vert}\right)\left(\hbar k_{\Vert}\over m\right)^2N{d^3k\over
\left(2\pi\right)^3}\cr
A_{\Vert}&=\int {8\pi ^2q^2\over\hbar}{R\over
\omega }\left|\vec{e}\cdot\vec{V}\right|^2\delta\left(\omega
-k_{\Vert}v_{\Vert}\right)\left(\hbar k_{\Vert}\over m\right){d^3k\over
\left(2\pi\right)^3},
\end{eqnarray}
where in the small gyroradius limit
$\vec{V}=\left(0,0,v_{\Vert}\right)$ and $\vec{e}$ is the wave
polarization vector. The number density of wave quanta is given by
$N$, $R$ is the ratio of electric energy to total energy in the wave, such that
$R\int N\hbar\omega d^3k/\left(2\pi\right)^3 = \delta E^2/8\pi$.
For lower hybrid waves\begin{equation} K^L=1+{\omega
_{pe}^2\over\Omega _e^2}\sin ^2\theta -{\omega _{pe}^2\over \omega
^2} +{\omega _{pi}^2\over kv_i^2}\left(1-\phi\left(\omega\over
\sqrt{2} kv_i\right)\right)\end{equation} where $\phi\left(x\right)$
is the usual plasma dispersion function and $v_i=\sqrt{k_BT_i/m_i}$
is the ion thermal velocity. We consider two possibilities. When
$\omega >> \sqrt{2}kv_i$, $\phi\simeq 1-k^2v_i^2/\omega ^2$ and
\begin{equation}
\omega ^2 = {\omega _{pi}^2+\omega _{pe}^2\cos ^2\theta\over 1+\omega _{pe}^2
\sin ^2\theta /\Omega _e^2},
\end{equation}
and when $\omega << \sqrt{2}kv_i$, $\phi\simeq\omega ^2/k^2v_i^2$, and
\begin{equation}
\omega ^2 ={\omega _{pe}^2\cos ^2\theta\over 1+\omega _{pe}^2
\sin ^2\theta /\Omega _e^2+\omega _{pi}^2/k^2v_i^2}.
\end{equation}
In either case, the ratio of electric to total energy in the wave
$R=\left(\omega\partial\Re K^L /\partial\omega\right)^{-1}\sim\Omega _e^2/2\omega _{pe}^2$ and $\left|\vec{e}
\cdot\vec{V}\right|^2=v_{\Vert}^2\cos ^2\theta$. For $\omega << \sqrt{2}kv_i$,
both $\omega$ and the growth rate $\gamma$ vary as $\cos ^2\theta$.
Consequently wave generation will occur
mainly at $\cos\theta =\omega _{pi}/\omega _{pe}$, beyond which lower hybrid waves are
progressively Landau damped by parallel propagating electrons. This then yields
\begin{eqnarray}
D_{\Vert\Vert}&=\pi {\partial\over\partial k_{\Vert}}\left(q\delta E\over
m_e\right)^2{\omega _{pi}^2\over \omega _{pe}^2v_{\Vert}}\cr
A_{\Vert}&={\omega _{pi}^2\Omega _e^4\over
8\pi \omega _{pe}^2n_ev_{te\perp}^2}
\end{eqnarray}
where $\partial /\partial k_{\Vert}\left(q\delta E/
m_e\right)^2$ is evaluated at $k_{\Vert}=\omega /v_{\Vert}$. For all cases of
interest, $A_{\Vert} << v_{\Vert}/\tau$ and so
\begin{equation}
f\sim \exp\left( -\int v_{\Vert}/\left[v_{t\Vert}^2+\tau D_{\Vert\Vert}\right]
dv_{\Vert}\right)=\left[v_{t\Vert}^2 +v_{\Vert}^2/2\kappa\right]^{-\kappa},
\end{equation}
where $\kappa =\left[4\pi\partial /\partial k_{\Vert}\left(q\delta E/
m_e\right)^2\times \pi\omega _{pi}^2n_e/3\omega _{pe}^6\ln\Lambda\right]^{-1}$.
We have assumed that $\partial /\partial k_{\Vert}\left(q\delta E/
m_e\right)^2$ is independent of $k_{\Vert}$ at $\omega /v_{\Vert}$ and hence that $D_{\Vert\Vert}
\propto 1/v_{\Vert}$ in the final step.

Where $\omega >> \sqrt{2}kv_i$, both $\omega$ and $\gamma$ remain non-zero
as $\cos\theta\rightarrow 0$, so wave generation occurs for all $\cos\theta$
in the range $-\omega _{pi}/\omega _{pe}\rightarrow\omega _{pi}/\omega _{pe}$.
In this case then
\begin{eqnarray}
D_{\Vert\Vert}&=\pi{\partial\over\partial k_{\Vert}}\left(q\delta E\over
m_e\right)^2{\omega ^2\over k^2v_{\Vert}^3}\cr
A_{\Vert}&={\omega ^2\Omega _e^4\over
8\pi k^2v_{\Vert}^2n_ev_{te\perp}^2}
\end{eqnarray}
and the electron distribution function integrates to a Maxwellian
\begin{eqnarray}
\nonumber f\left(v_{\Vert}\right)&\propto\exp\left(-v_{\Vert}^2/2/\left(v_{t\Vert}^2
+\left(\pi n_e\omega ^2/6\omega _{pe}^4\ln\Lambda k^2\right)4\pi
\partial/\partial k_{\Vert}\left(q\delta E/m_e\right)^2\right)\right)\cr &=
\exp\left(-v_{\Vert}^2/2/\left( v_{t\Vert}^2
+\left(\omega ^2/k^2v_{\Vert}^2\right)\left(\tau _s\pi/v_{\Vert}\right)
\partial/\partial k_{\Vert}\left(q\delta E/m_e\right)^2\right)\right).
\end{eqnarray}
We also estimate
$A_{\Vert}\simeq 10^{11}/n_e << v_{\Vert}/\tau _s$ as required with $B=10$G.

\section{Atomic Data Updates}
We have also updated the atomic data for dielectronic recombination of K-shell,
and L-shell ions. Recombination from H- to He-like and from He- to Li-like are
taken from \citet{dasgupta04}. The successive isoelectronic sequences Li-,
Be-, B-, C-, N-, O-, and F-like are taken from \citet{colgan04},
\citet{colgan03}, \citet{altun04}, \citet{zatsarinny04a}, \citet{mitnik04},
\citet{zatsarinny03}, and \citet{gu03} respectively. Additionally dielelectronic
recombination from Ne- to Na-like and from Na- to Mg-like are taken from
\citet{zatsarinny04b} and \citet{gu04}.

\clearpage

\begin{table}[t]
\begin{center}
\caption{O$^{+6}$/O$^{+7}$ Abundance Ratios}
\begin{tabular}{lrrrr}
\tableline\tableline
$v_{start}$ (km s$^{-1}$) & $\gamma ^{\prime}$& $\kappa _{min}$& O$^{+6}$/O$^{+7}$&
O$^{+6}$/O$^{+7}_{\kappa\rightarrow\infty}$\\
\tableline
5& 0.005& 4.00& 310& 360\\
& 0.01& 3.40& 247& 205\\
&  0.02& 3.07& 156& 79\\
&  0.03& 2.95& 115& 50\\
&  0.04& 2.89& 95& 38\\
&  0.05& 2.84& 84& 32\\
&  0.06& 2.82& 74& 28\\
&  0.08& 2.78& 65& 24\\
&  0.1&  2.75& 58& 22\\
\\
10& 0.005& 4.78& 278& 328\\
 & 0.01& 3.78& 192& 145\\
 &  0.02& 3.29& 89& 46\\
 & 0.03& 3.12& 59& 27\\
 & 0.04& 3.03& 46& 21\\
 & 0.05& 2.97& 38& 17\\
 & 0.06& 2.93& 34& 15\\
 & 0.08& 2.87& 29& 13\\
 & 0.1& 2.84& 26& 11\\
 \\
20& 0.005& 5.92& 221& 278\\
 & 0.01& 4.38& 118& 99\\
 & 0.02& 3.62&  46& 26\\
 & 0.03& 3.36& 27& 15\\
 & 0.04& 3.23& 21& 11\\
 & 0.05& 3.14& 17& 9\\
 & 0.06& 3.08& 15& 9\\
 & 0.08& 3.00& 12& 7\\
 & 0.1& 2.96& 11& 6\\
 \\
Geiss et al. (1995) & & & & 30\\
Ko et al. (1997) & & & & 32\\
 \tableline
\end{tabular}
\end{center}
\label{tab1}
\end{table}

\begin{table}[t]
\begin{center}
\caption{Fe Charge State Fractions}
\begin{tabular}{lrrrrrrrrrr}
\tableline\tableline
$v_{start}$ (km s$^{-1}$) & $\gamma ^{\prime}$& Fe$^{+7}$& Fe$^{+8}$& Fe$^{+9}$& Fe$^{+10}$& Fe$^{+11}$&
Fe$^{+12}$& Fe$^{+13}$& Fe$^{+14}$\\
\tableline
& 0.0  & 0.21 & 0.63& 0.094& 0.025& 0.0044& 0.0004\\
& 0.005& 0.041& 0.60& 0.29& 0.059& 0.006& 0.0004& \\
& 0.007& 0.019& 0.49& 0.37& 0.11& 0.015& 0.001&  \\
& 0.01& 0.007& 0.34& 0.42& 0.19& 0.039& 0.005& 0.0002& \\
& 0.015& 0.003& 0.20& 0.40& 0.28& 0.093& 0.017& 0.001& \\
&  0.02& 0.0012& 0.14& 0.35& 0.33& 0.14& 0.035& 0.004& 0.0003\\
& 0.025& 0.0007& 0.10& 0.31& 0.35& 0.18& 0.054& 0.007& 0.0007\\
&  0.03& 0.0005& 0.076& 0.28& 0.35& 0.21& 0.073& 0.012& 0.001\\
&  0.04& 0.0003& 0.052& 0.23& 0.35& 0.25& 0.10& 0.021& 0.003\\
&  0.05& 0.0002& 0.040& 0.20& 0.34& 0.27& 0.13& 0.029& 0.004\\
&  0.06& 0.0001& 0.032& 0.17& 0.32& 0.28& 0.15& 0.036& 0.006\\
&  0.08&       & 0.024& 0.15& 0.31& 0.30& 0.17& 0.047& 0.008\\
\\
10& 0.0& 0.17& 0.62& 0.13& 0.043& 0.009& \\
 & 0.005& 0.020& 0.47& 0.37& 0.12& 0.017& 0.0013& \\
 & 0.007& 0.0075& 0.32& 0.42& 0.21& 0.047& 0.006&  \\
 &  0.01& 0.0023& 0.17& 0.38& 0.31& 0.11& 0.022& 0.002& \\
 & 0.015& 0.0005& 0.070& 0.26& 0.36& 0.22& 0.077& 0.012& 0.001\\
 &  0.02& & 0.035& 0.18& 0.33& 0.28& 0.14& 0.030& 0.004\\
 & 0.025& & 0.020& 0.13& 0.29& 0.31& 0.19& 0.052& 0.009\\
 & 0.03& & 0.013& 0.097& 0.26& 0.31& 0.23& 0.074& 0.015\\
 & 0.04& & 0.006& 0.062& 0.20& 0.31& 0.28& 0.11& 0.029\\
 & 0.05& & 0.004& 0.044& 0.17& 0.29& 0.30& 0.14& 0.043\\
 & 0.06& & 0.003& 0.034& 0.14& 0.27& 0.32& 0.17& 0.055\\
 & 0.08& & 0.002& 0.023& 0.11& 0.24& 0.33& 0.20& 0.074\\
 \\
20& 0.0& 0.14& 0.58& 0.18& 0.074& 0.020& 0.003& \\
 & 0.005& 0.003& 0.13& 0.31& 0.34& 0.17& 0.047& 0.007& 0.0002\\
 & 0.007& 0.0008& 0.057& 0.22& 0.35& 0.26& 0.10& 0.016& 0.0013\\
 & 0.01& 0.0001& 0.016& 0.10& 0.27& 0.32& 0.22& 0.062& 0.010\\
 & 0.015& & 0.0039& 0.041& 0.16& 0.29& 0.32& 0.14& 0.038\\
 & 0.02& & 0.0013& 0.019& 0.099& 0.24& 0.34& 0.21& 0.077\\
 & 0.025& & 0.002& 0.023& 0.11& 0.25& 0.34& 0.19& 0.069\\
 & 0.03& & 0.0008& 0.013& 0.076& 0.20& 0.34& 0.24& 0.10\\
 & 0.04& & 0.0002& 0.005& 0.042& 0.14& 0.31& 0.28& 0.16\\
 & 0.05& & 0.0001& 0.003& 0.026& 0.10& 0.28& 0.30& 0.20\\
 & 0.06& & & 0.002& 0.018& 0.082& 0.25& 0.30& 0.22\\
 & 0.08& & & 0.001& 0.011& 0.057& 0.21& 0.29& 0.25\\

 \\
 \tableline
\end{tabular}
\end{center}
\label{tab2}
\end{table}

\begin{table}[t]
\begin{center}
\caption{Minor Ion Charge State Fractions, $v_{start}=10$ km s$^{-1}$}
\begin{tabular}{lrrrrrrrrrr}
\tableline\tableline
element  & $\gamma ^{\prime}$& +5 & +6 & +7& +8 &+9& +10& +11&
+12\\
\tableline
Ne & 0.01 & 0.0017&  0.032& 0.12& 0.84& 0.0003\\
   & 0.015& 0.0005& 0.024& 0.12& 0.85& 0.001\\
   & 0.025& 0.0002& 0.017& 0.12& 0.86& 0.004\\
\\
Mg & 0.01& 0.008& 0.12& 0.40& 0.35& 0.084& 0.038\\
   & 0.015& 0.004& 0.089& 0.38& 0.40& 0.095& 0.036\\
   & 0.025& 0.002& 0.058& 0.33& 0.45& 0.123& 0.036\\
   \\
Si & 0.01& 0.004& 0.12& 0.47& 0.34& 0.070& 0.004\\
   & 0.015& 0.001& 0.058& 0.38& 0.42& 0.13& 0.013\\
   & 0.025& 0.0002& 0.022& 0.26& 0.45& 0.23& 0.037& 0.002\\
   \\
S & 0.01& 0.0004& 0.089& 0.42& 0.39& 0.095& 0.007& 0.0002\\
  & 0.015& 0.0001& 0.047& 0.33& 0.44& 0.16& 0.020& 0.0009\\
  & 0.025& & 0.017& 0.21& 0.44& 0.28& 0.058& 0.005\\
  \\
Ar & 0.01& & & 0.006& 0.69& 0.28& 0.032& 0.0013\\
  & 0.015& & & 0.003& 0.59& 0.34& 0.058& 0.004\\
  & 0.025& & & 0.0013& 0.43& 0.43& 0.13& 0.015& 0.0007\\
\\
Ca & 0.01& & & 0.005& 0.034& 0.10& 0.81& 0.048& 0.0008\\
   & 0.015& & & 0.002 & 0.016& 0.084& 0.80& 0.095& 0.004\\
   & 0.025& & & 0.0003& 0.004& 0.045& 0.75& 0.18& 0.015\\
 \tableline
\end{tabular}
\end{center}
\label{tab3}
\end{table}

\clearpage

\begin{figure}
\epsscale{1.0} \plotone{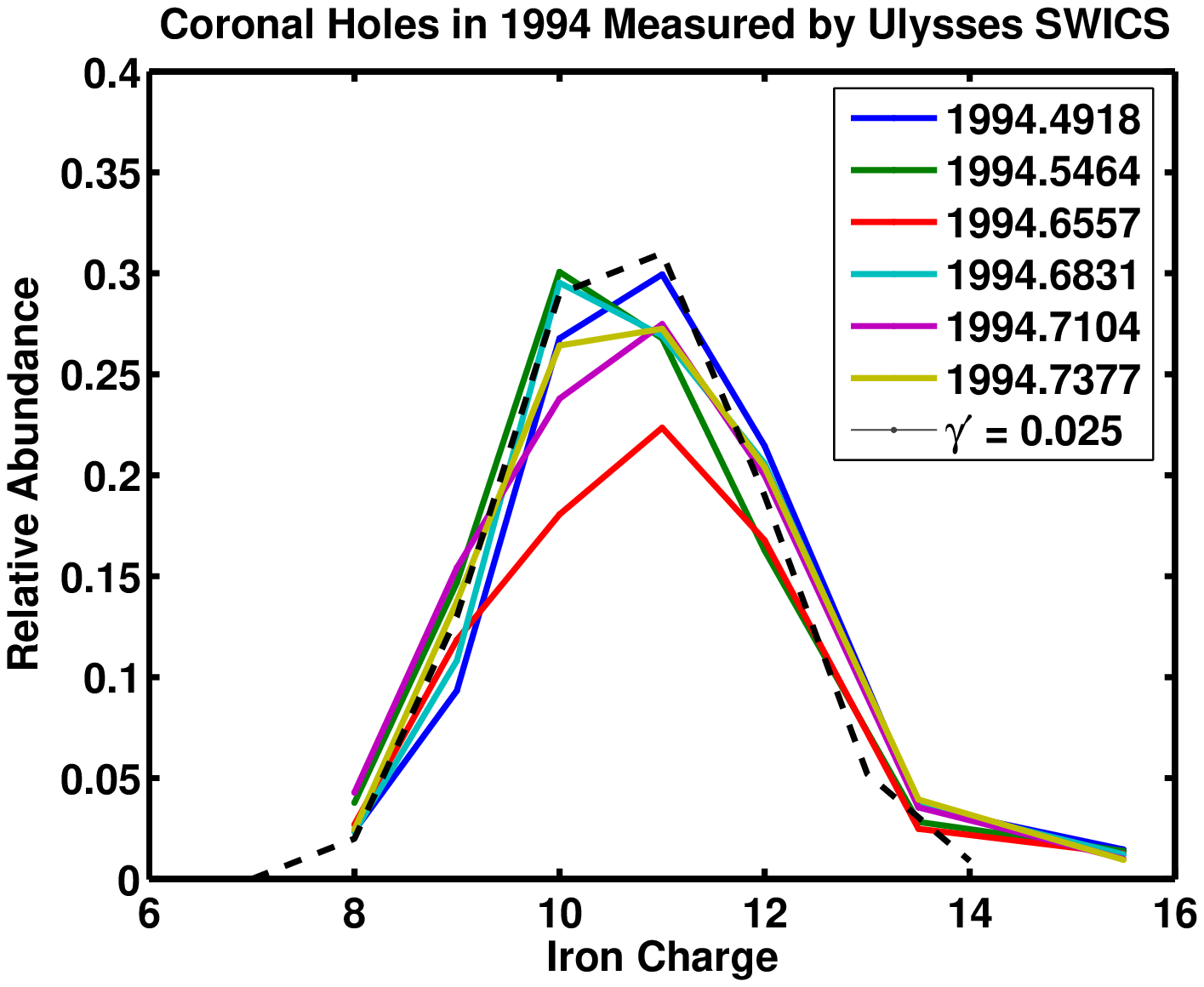}
\figcaption[f2.eps]{Observed (colored solid lines) and modeled (dashed line) Fe charge states
in fast wind from polar coronal holes observed at solar minimum by Ulysses SWICS in 1994.\label{fig2}}
\end{figure}

\begin{figure}
\epsscale{1.0} \plotone{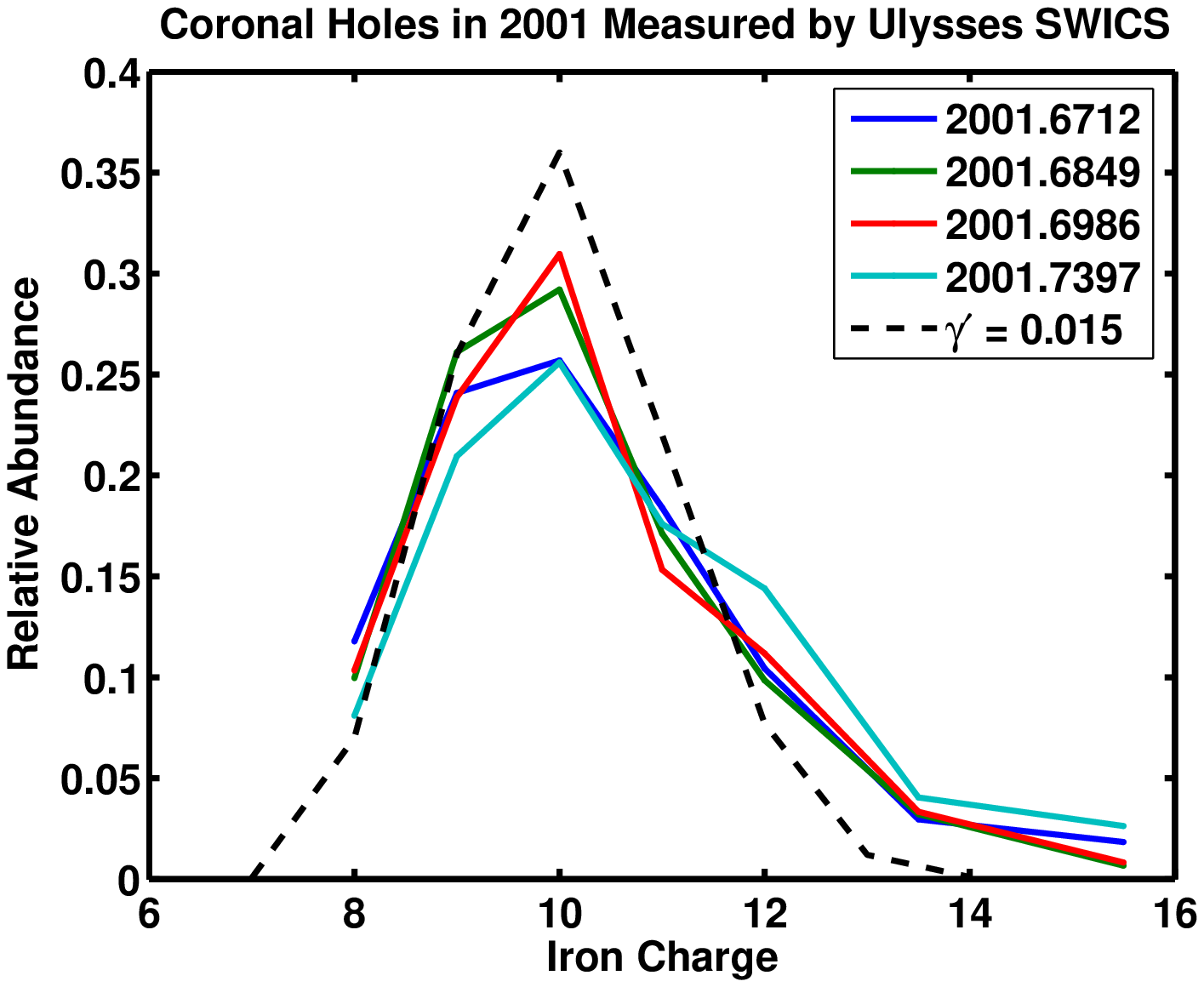}
\figcaption[f3.eps]{Observed (colored solid lines) and modeled (dashed line) Fe charge states
in fast wind from polar coronal holes observed at solar maximum by Ulysses SWICS in 2001.\label{fig3}}
\end{figure}

\begin{figure}
\epsscale{1.0} \plotone{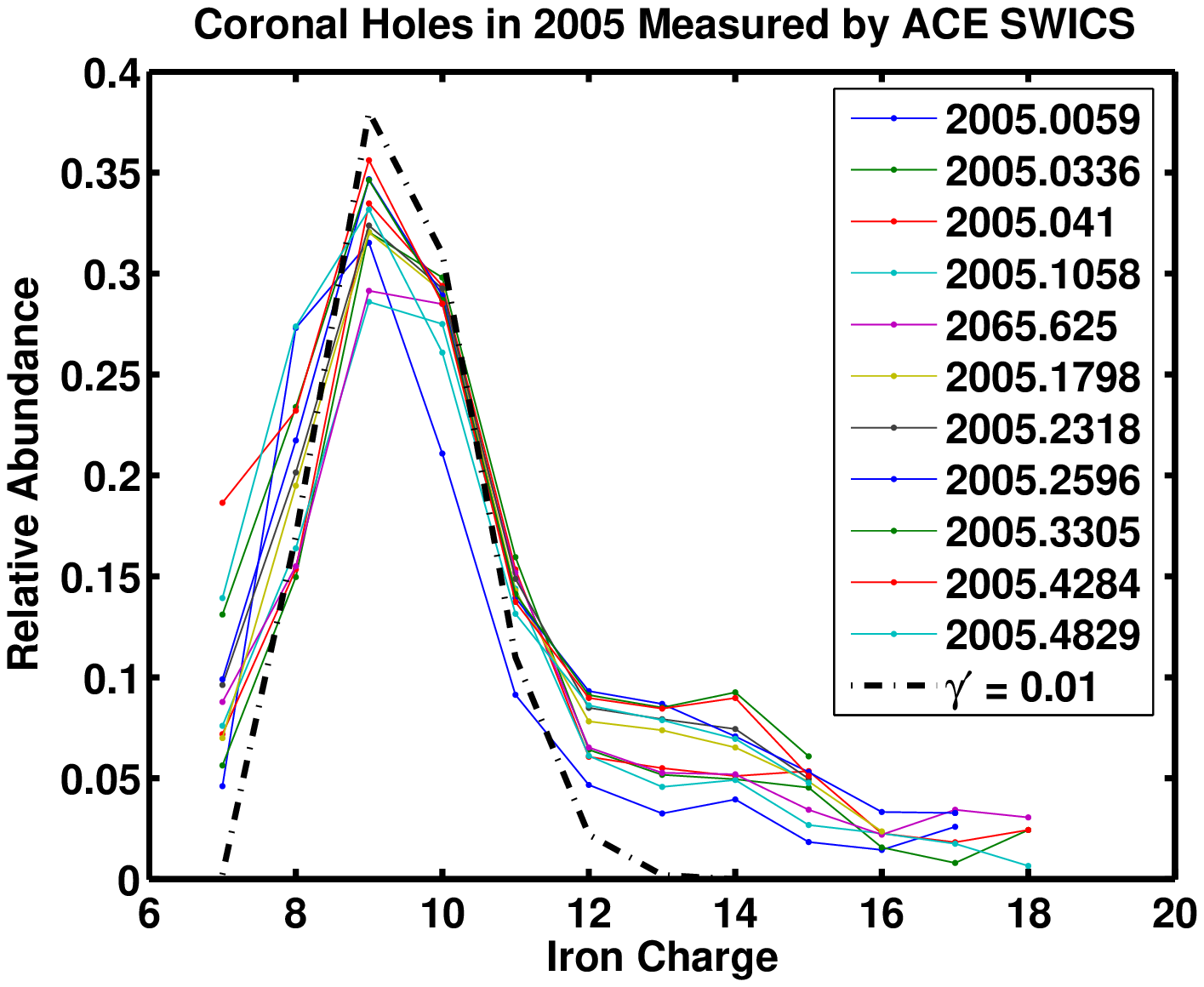}
\figcaption[f4.eps]{Observed (colored solid lines) and modeled (dashed line) Fe charge states
in fast wind peaking at Q$_{Fe}$ = 9 from equatorial coronal holes observed by ACE in 2005. The average charge states are
lower still than the solar maximum coronal holes, and the distributions are generally broader.
The $\alpha$ particle wind speed is also the lowest in these cases.\label{fig4}}
\end{figure}

\begin{figure}
\epsscale{1.0} \plotone{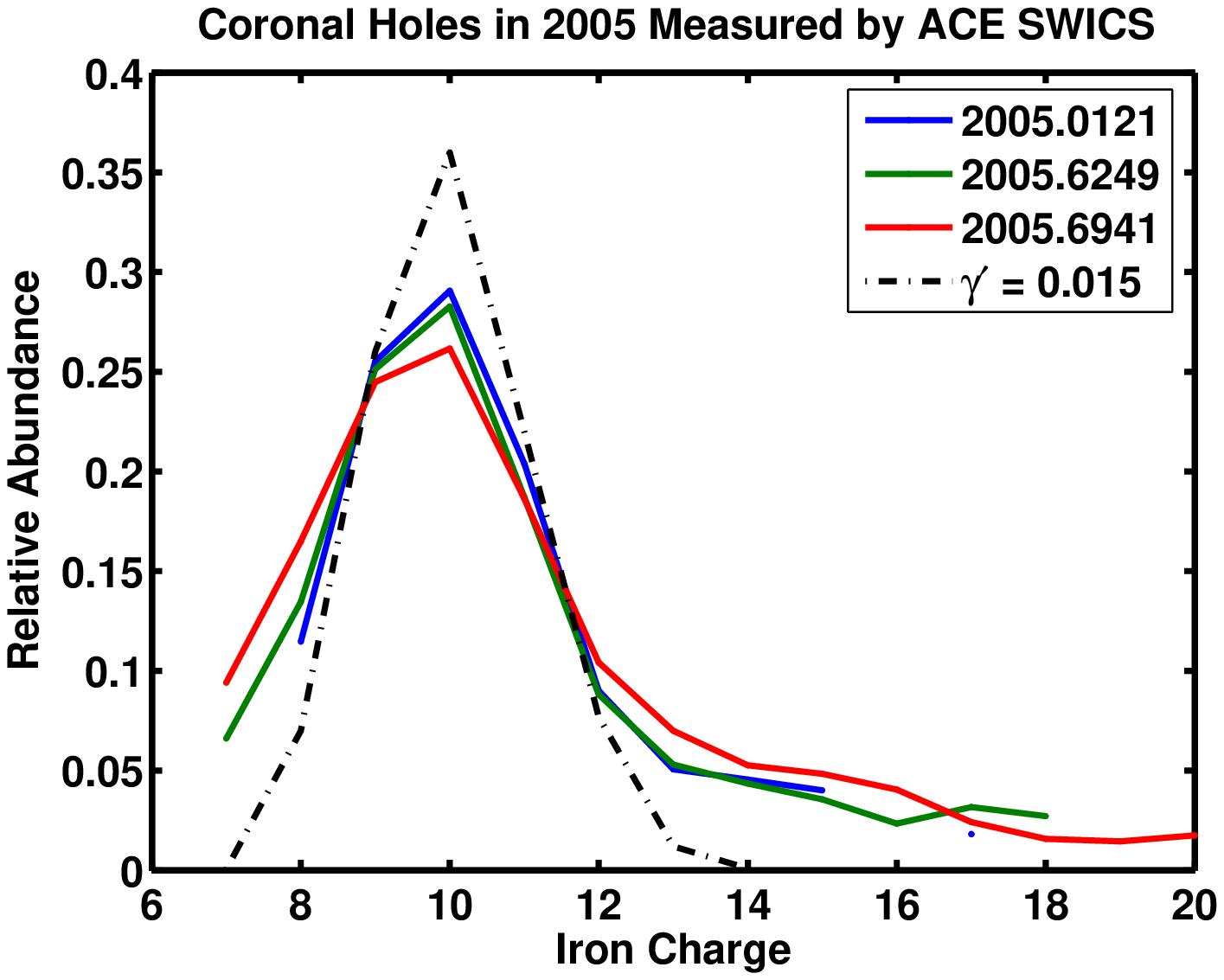}
\figcaption[f5.eps]{Observed (colored solid lines) and modeled (dashed line) Fe charge states
in fast wind peaking at Q$_{Fe}$ = 10 from equatorial coronal holes observed by ACE in 2005. The average charge states are
lower still than the solar maximum coronal holes, and the distributions are generally broader.
The $\alpha$ particle wind speed is also the lowest in these cases.\label{fig5}}
\end{figure}

\end{document}